\documentclass[aps,prd,preprint,showkeys,superscriptaddress]{revtex4}

\usepackage{latexsym}
\usepackage{amsmath,amssymb}

\usepackage{filecontents}   

\begin{document}


\title{Entropy and temperatures of Nariai black hole}

\author{Myungseok Eune} %
\email[]{younms@sogang.ac.kr} %
\affiliation{Research Institute for Basic Science, Sogang University,
  Seoul, 121-742, Korea} %

\author{Wontae Kim} %
\email[]{wtkim@sogang.ac.kr} %
\affiliation{Research Institute for Basic Science, Sogang University,
  Seoul, 121-742, Korea} %
\affiliation{Center for Quantum Spacetime, Sogang University, Seoul
  121-742, Korea} %
\affiliation{Department of Physics, Sogang University, Seoul 121-742,
  Korea} %

\date{\today}

\begin{abstract}
  The statistical entropy of the Nariai black hole in a thermal
  equilibrium is calculated by using the brick-wall method. Even if
  the temperature depends on the choice of the time-like Killing
  vector, the entropy can be written by the ordinary area law which
  agrees with the Wald entropy.  We discuss some physical consequences
  of this result and the properties of the temperatures.
\end{abstract}


\maketitle

\section{Introduction}
\label{sec:intro}

It has been claimed that the entropy of a black hole is proportional
to the surface area at the event horizon~\cite{Bekenstein:1973ur}, and
then the Schwarzschild black hole has been studied through the quantum
field theoretic calculation~\cite{Hawking:1974sw}.  One of the
convenient methods to get the entropy is to use the brick-wall method,
which gives the statistical entropy satisfying the area law of the
black hole~\cite{'tHooft:1984re}.  Then, there have been extensive
applications to various black holes~\cite{Mann:1990fk,Ghosh:1994wb,
  Ghosh:1994mm,Solodukhin:1996vx,Ho:1998du,Li:2000rk,Winstanley:2000in,
  Li:2002xb,Kim:2005pb,Yoon:2007aj,Kim:2007if,Kim:2007nh,Ren:2008rv,
  Liu:2008bb,Sarkar:2007uz,Ghosh:2009xg,Kim:2008bf,ChangYoung:2008pw,
  Eune:2010kx,Kay:2011np,Son:2011ii,Zhao:2011zzc,Cao:2012zzb}.
In fact, another way to obtain the entropy is to regard the entropy of
black holes as the conserved Noether charge corresponding to the
symmetry of time translation~\cite{Wald:1993nt}. For the Einstein
gravity, the Wald entropy is always given by the $A_H/(4G)$, where
$A_H$ and $G$ are the surface area at the event horizon and the
Newton's gravitational constant, respectively. Actually, there are
many extended studies for the entropy as the Noether charge in the
general theory of gravity including the higher power of the
curvature~\cite{Iyer:1995kg, Jacobson:1993vj, Jacobson:1993xs,
  LopesCardoso:1999cv, Parikh:2009qs, Parikh:2009js}.

The fact that the cosmological constant seems to be positive in our
universe deserves to study the Schwarzschild black hole on the de
Sitter background, which can be easily realized in the form of the
Schwarzschild-de Sitter (SdS) spacetime.  It has the black hole
horizon and the cosmological horizon, and the observer lives between
them.  In this spacetime, the temperature of the black hole is
different from the temperature due to the cosmological horizon when $0
< M < 1/(3\sqrt{\Lambda})$, where $M$ and $\Lambda$ are the mass of
the black hole and the cosmological constant, respectively. Therefore,
we are in trouble to study the thermodynamics of the system since it
is not in thermally equilibrium due to the different
temperatures. Nevertheless, there are several studies for the entropy
through the improved brick-wall method for the SdS black
hole~\cite{Zhao:2001pf, Wu:2001uy, Ding:2007ur} and the Kerr-de Sitter
black hole~\cite{Wu:2003qc}. In order to avoid the difficulty due to
the nonequilibrium state of the SdS black hole, they have considered
two thin-layers near the black hole horizon and the cosmological
horizon, and then calculated the entropy for each thin layer.

On the other hand, for the special limit of $M = 1 /
(3\sqrt{\Lambda})$ in the SdS spacetime, the two horizons are
coincident in the Schwarzschild coordinate.  However, the Nariai
metric is obtained through the coordinate transformation to avoid the
coordinate singularity, where the two horizons are still
separated~\cite{Ginsparg:1982rs, Bousso:1996au, Bousso:1997wi,
  Dias:2003up}. The Nariai spacetime is in thermally equilibrium since
the black hole and the cosmological horizon give the same
temperatures.  Thus, we can treat the whole Nariai spacetime as one
thermodynamical system.  However, even in spite of thermodynamic
equilibrium, there few thermodynamic studies. Moreover, one can define
two kinds of temperatures for the Nariai black hole: the
Bousso-Hawking temperature and the Hawking temperature since there
exist two different normalizations of timelike Killing
vectors~\cite{Bousso:1996au}.

In this paper, we would like to study the statistical entropy of the
Nariai black hole by using the brick-wall method. In
section~\ref{sec:NariaiBH}, we introduce the SdS spacetime and the
Nariai spacetime, and define two kinds of temperatures based on the
different normalizations of the Killing vectors.  We will also apply
the Wald formula tor the Nariai black hole in order to get the entropy
without resort to normalizations of the Killing vector.  In
section~\ref{sec:brick-wall}, the entropy will be calculated by using
the brick-wall method. Although the energy and the temperature depend
on the normalization of the time-like Killing vector, the
normalization-independent statistical entropy can be obtained, which
is compatible with the Wald entropy.  Finally, summary and discussion
are given in section~\ref{sec:discus}.

\section{Temperatures and Wald entropy in Nariai black hole}
\label{sec:NariaiBH}

Let us start with the four-dimensional Einstein-Hilbert action with
the cosmological constant $\Lambda$, which is given by
\begin{align}
  I = \frac{1}{16\pi G} \int d^4 x \sqrt{-g} (R -
  2\Lambda). \label{action:EH}
\end{align}
The equation of motion obtained from the action~\eqref{action:EH}
becomes
\begin{align}
  R_{\mu\nu} - \frac12 g_{\mu\nu} R + \Lambda g_{\mu\nu} =
  0. \label{eom}
\end{align}
The static and spherically symmetric solution of Eq.~\eqref{eom} is
written as
\begin{align}
  ds^2 = - f(r) dt^2 + \frac{dr^2}{f(r)} + r^2
  d\Omega_2^2, \label{metric:SdS}
\end{align}
with
\begin{align}
  f(r) = 1- \frac{2M}{r} - \frac13 \Lambda
  r^2. \label{metric:function:SdS}
\end{align}
Hereafter, we will consider only the Schwarzschild-de Sitter spacetime
with $\Lambda >0$.  For $0<M \le 1/(3\sqrt{\Lambda})$, it has two
horizons of the black hole horizon $r_b$ and the cosmological horizon
$r_c$. In this case, the metric function~\eqref{metric:function:SdS}
can be neatly written as $f(r) = (1-r_b/r)[1 - (\Lambda/3)(r^2 + r_b r
+ r_b^2)] = (r-r_b) (r_c - r) (r + r_b + r_c)/[r (r_b^2 + r_b r_c +
r_c^2)]$. For $M=0$, it has only the cosmological horizon with $r_c =
\sqrt{3/\Lambda}$.

The symmetry of time translation in the SdS spacetime can be described
by a timelike Killing vector, which is written as
\begin{align}
  \label{killing:t}
  \xi = \gamma_t \frac{\partial}{\partial t},
\end{align}
where $\gamma_t$ is a normalization constant. In the standard
normalization, $\gamma_t$ is obtained from the condition to satisfy
$\xi^\mu \xi_\mu = -1$ at the asymptotically flat Minkowski spacetime.
For instance, its value usually becomes $\gamma_t = 1$ for a
Schwarzschild metric.  In the SdS spacetime, there is no
asymptotically flat region, so that we should consider the reference
point $r_g$ where the gravitational acceleration vanishes due to the
balance between the forces of the black hole by the mass and the
cosmological horizon by the cosmological constant. Thus, we can choose
the normalization constant in Eq.~\eqref{killing:t} to satisfy
$\xi^\mu \xi_\mu = -1$ at that reference point $r_g$, which yields
\begin{align}
  \gamma_t = \frac{1}{\sqrt{f(r_g)}}, \label{gamma:t:rg}
\end{align}
where the reference point can be found from $f'(r_g) = 0$ and is
explicitly given by $r_g = \left(3M / \Lambda \right)^{1/3}$.  Now,
the surface gravities $\kappa_b$ and $\kappa_c$ on the black hole
horizon and the cosmological horizon are written as
\begin{align}
  \kappa_{b,c} = \lim_{r \to r_{b,c}} \sqrt{\frac{\xi^\mu \nabla_\mu
      \xi_\nu \xi^\rho\nabla_\rho \xi^\nu}{-\xi^2}}, \label{def:kappa}
\end{align}
respectively. 
Then, the temperatures along with the normalization~\eqref{gamma:t:rg} are
calculated as
\begin{align}
  T_{\rm BH}^{b,c} = \frac{\kappa_{b,c}}{2\pi} =
  \frac{f'(r_{b,c})}{4\pi\sqrt{f(r_g)}}, \label{T:b,c:rg}
\end{align}
which are called the Bousso-Hawking
temperatures~\cite{Bousso:1996au}. This temperature can be also
obtained from $\tilde\xi = \partial / \partial \tilde{t}$ when the
time is rescaled as $\tilde{t} = t \sqrt{f(r_g)}$, where $\xi^\mu
\xi_\mu = -1$ is satisfied at $r=r_g$.

On the other hand, in the Euclidean geometry, the Hawking temperature
agrees with the inverse of the period of the Euclidean time to avoid a
conical singularity at the horizon. Setting the Euclidean time $\tau$
to $\tau = i t$, the Euclidean line element of Eq.~\eqref{metric:SdS}
is written as
\begin{align}
  ds_E^2 = f(r) d\tau^2 + \frac{dr^2}{f(r)} + r^2
  d\Omega_2^2. \label{dS:Euclidean}
\end{align}
From Eq.~\eqref{dS:Euclidean}, the Hawking temperatures for the black
hole horizon and the cosmological horizon become
\begin{align}
  T_{\rm H}^{b,c} = \beta_{\rm H}^{-1} =
  \frac{f'(r_{b,c})}{4\pi}, \label{T:H:standard}
\end{align}
respectively, which agree with the temperatures obtained from the
Killing vector~\eqref{killing:t} with the normalization constant
$\gamma_t = 1$.  Note that the Hawking
temperature~\eqref{T:H:standard} is definitely different from the
Bousso-Hawking temperature~\eqref{T:b,c:rg}. For the scaled Euclidean
time given by $\tilde{\tau} = i \tilde{t}$, the Bousso-Hawking
temperatures are obtained.

Similar argument can be done for the Nariai black hole by taking the
limit of $M = 1/(3\sqrt{\Lambda})$ in Eq.~\eqref{metric:function:SdS}
so that the two horizons are coincident in the Schwarzschild
coordinate. In this degenerate case with $r_b = r_c$, the
metric~\eqref{metric:SdS} should be transformed to an appropriate
coordinate system because it has the coordinate singularity and
becomes inappropriate. Near the degenerate case, the mass can be
written as~\cite{Ginsparg:1982rs, Bousso:1996au, Bousso:1997wi}
\begin{align}
  9M^2 \Lambda = 1 - 3 \epsilon^2, \quad 0 \le \epsilon \ll
  1, \label{def:epsilon}
\end{align}
where the degenerate case can be obtained by taking $\epsilon =0$.
One can define the new time and the radial coordinate $\psi$ and
$\chi$ by
\begin{align}
  t = \frac{1}{\epsilon\sqrt{\Lambda}} \psi, \quad r =
  \frac{1}{\sqrt{\Lambda}} \left(1 - \epsilon \cos\chi - \frac16
    \epsilon^2 \right). \label{def:psi,chi}
\end{align}
In terms of the new coordinates~\eqref{def:psi,chi}, the line
element~\eqref{metric:SdS} is written in the form of
\begin{align}
  ds^2 = \frac{1}{\Lambda} \left[- \left(1 + \frac23 \epsilon
      \cos\chi\right) \sin^2\chi d\psi^2 + \left(1- \frac23 \epsilon
      \cos\chi\right) d\chi^2 + (1-2\epsilon \cos\chi)
    d\Omega_2^2\right], \label{metric:near_Nariai}
\end{align}
up to the first order in $\epsilon$. For the case of $\epsilon =0$,
Eq.~\eqref{metric:near_Nariai} is called the Nariai metric, which is
given by
\begin{align}
  ds^2 = \frac{1}{\Lambda} \left(- \sin^2\chi d\psi^2 + d\chi^2 +
    d\Omega_2^2\right). \label{metric:Nariai}
\end{align}
In this coordinate system, the back hole horizon and the cosmological
horizon correspond to $\chi =0$ and $\chi=\pi$, respectively, where
the proper distance between the two horizons is given by
$\pi/\sqrt{\Lambda}$ which is not zero.  From now on, we will study
this Nariai black hole which is actually real geometry to describe
thermal equilibrium since the horizon temperature is the same with the
cosmological temperature. However, there are two kinds of temperatures
depending on the definitions of the normalization of the Killing
vector.

With Eqs.~\eqref{def:epsilon} and~\eqref{def:psi,chi}, the Killing
vector~\eqref{killing:t} becomes
\begin{align}
  \xi = \sqrt{\Lambda} \frac{\partial}{\partial
    \psi}, \label{killing:psi}
\end{align}
to the leading order in $\epsilon$. Using Eq.~\eqref{def:kappa}, the
Bousso-Hawking temperature is calculated as
\begin{align}
  T_{\rm BH}^{b,c} = \frac{\sqrt{\Lambda}}{2\pi}. \label{T:BH:Nariai}
\end{align}
This can be also obtained from the Killing vector $\tilde\xi
= \partial / \partial \tilde\psi$ at the coordinate system with the
rescaled time $\tilde\psi = \psi/\sqrt{\Lambda}$.  As expected, the
temperature of the black hole horizon is the same with that of the
cosmological horizon.  On the other hand, one can also get the Hawking
temperature from the Euclidean metric~\eqref{metric:Nariai} by setting
the Euclidean time as $\psi_E = i\psi$. Then, the Euclidean Nariai
metric can be written as
\begin{align}
  ds_E^2 = \frac{1}{\Lambda} \left(\sin^2\chi\, d\psi_E^2 + d\chi^2 +
    d\Omega_2^2 \right). \label{metric:Nariai:Euclidean}
\end{align}
In order to avoid a conical singularity at the two horizons, the
period of the Euclidean time for the black hole horizon or the
cosmological horizon are chosen as $2\pi$, respectively.  Then the
Hawking temperatures are given by
\begin{align}
  T_H^{b,c} = \beta_H^{-1} = \frac{1}{2\pi}, \label{T:H:Nariai}
\end{align}
which corresponds to the surface gravity obtained from the Killing
vector $\partial/\partial\psi$ using Eq.~\eqref{def:kappa}.  It is
interesting to note that the Hawking temperature is constant as long
as the Nariai condition $M = 1/(3\sqrt{\Lambda})$ is met. Moreover, it
can be easily checked that the Bousso-Hawking
temperature~\eqref{T:BH:Nariai} is obtained from the condition to
avoid a conical singularity at the horizons for the scaled Euclidean
time $\tilde{\psi}_E = \psi_E / \sqrt{\Lambda}$.

In order to find the Wald entropy of the Nariai spacetime, one should
consider a diffeomorphism invariance with the Killing vector $\xi^\mu$
which is associated with the conservation law of $\nabla_\mu J^\mu =
0$~\cite{Wald:1993nt, Iyer:1995kg, Jacobson:1993vj, Jacobson:1993xs,
  LopesCardoso:1999cv, Parikh:2009qs}, for which the Noether potential
$J^{\mu\nu}$ can be defined by $J^\mu = \nabla_\nu J^{\mu\nu} $. If a
Lagrangian is written in the form of $\mathcal{L} = \mathcal{L}
(g_{\mu\nu}, R_{\mu\nu\rho\sigma})$, then the Noether potential is
given by~\cite{LopesCardoso:1999cv, Parikh:2009qs}
\begin{align}
  J^{\mu\nu} = - 2 \Theta^{\mu\nu\rho\sigma} \nabla_\rho \xi_\sigma +
  4 \nabla_\rho \Theta^{\mu\nu\rho\sigma}
  \xi_\sigma, \label{def:Noether_potential}
\end{align}
where 
\begin{align}
  \Theta^{\mu\nu\rho\sigma} = \frac{\partial \mathcal{L}}{\partial
    R_{\mu\nu\rho\sigma}}. \label{def:Theta}
\end{align}
For a timelike Killing vector, the Wald entropy~\cite{Wald:1993nt} is
expressed by
\begin{align}
  S = \frac{2\pi}{\kappa} \int_{\Sigma}  d^2 x \sqrt{h}\,
  \epsilon_{\mu\nu} J^{\mu\nu}, \label{formula:Wald_entropy}
\end{align}
where $\kappa$ and $h_{\mu\nu}$ are the surface gravity and the
induced metric on the hypersurface $\Sigma$ of a horizon,
respectively. And $\epsilon_{\mu\nu}$ is defined by
\begin{align}
  \epsilon_{\mu\nu} = \frac12 (n_\mu u_\nu - n_\nu
  u_\mu), \label{def:epsilon_ab}
\end{align}
where $n^\mu$ is the outward unit normal vector of $\Sigma$. The
proper velocity $u^\mu$ of a fiducial observer moving along the orbit
of $\xi^\mu$ is given by $u^\mu = \alpha^{-1} \xi^\mu$ with $\alpha
\equiv \sqrt{-\xi^\mu \xi_\mu}$.

For the Nariai metric~\eqref{metric:Nariai}, the Killing vector is
given by
\begin{align}
  \xi = \gamma \frac{\partial}{\partial \psi}, \label{Nariai:Killing}
\end{align}
where $\gamma$ is a normalization constant, which will be not
specified in this section. From the norm of the Killing vector, we
obtain $\alpha=\gamma \sin\chi / \sqrt{\Lambda}$ and $u_\mu =
\xi_\mu/\alpha = - \delta_\mu^\psi \sin\chi/ \sqrt{\Lambda}$. The
outward unit normal vectors of the black hole horizon and the
cosmological horizon are calculated as $n_\mu = (1 / \sqrt{\Lambda})
\delta_\mu^\chi$ and $n_\mu = - (1 / \sqrt{\Lambda}) \delta_\mu^\chi$,
respectively. Then, the nonzero components of
Eq.~\eqref{def:epsilon_ab} are $\epsilon_{\psi\chi} = -
\epsilon_{\chi\psi} = \pm \sin\chi / (2\Lambda)$, where the upper sign
and the lower sign correspond to the black hole horizon and the
cosmological horizon, respectively. Now, for the
action~\eqref{action:EH}, we obtain
\begin{align}
  \Theta^{\mu\nu\rho\sigma} = \frac{1}{32\pi G}(g^{\mu\rho} g^{\nu\sigma} -
  g^{\mu\sigma} g^{\nu\rho}), \label{Theta:value:R}
\end{align}
which leads to
\begin{align}
  \epsilon_{\mu\nu} J^{\mu\nu} = \pm \frac{\gamma}{8\pi G}
  \cos\chi. \label{eJ:value}
\end{align}
Inserting Eq.~\eqref{Nariai:Killing} into Eq.~\eqref{def:kappa}, we
can obtain $\kappa_{b,c} = \gamma$. Then, from
Eq.~\eqref{formula:Wald_entropy}, the Wald entropy is given by
\begin{align}
  S = \frac{1}{4G} \left( \int_{\Sigma_{\chi=0}} d^2 x \sqrt{h}
    \cos\chi - \int_{\Sigma_{\chi=\pi}} d^2x \sqrt{h} \cos\chi \right)
  = \frac{A_b + A_c}{4G}, \label{S:Wald:bc}
\end{align}
where $A_b$ and $A_c$ are the areas of the black hole horizon and the
cosmological horizon, respectively.  The total area given by the two
horizons becomes $A = 8\pi / \Lambda$ since $A_b = A_c =
4\pi/\Lambda$. Eventually, the entropy~\eqref{S:Wald:bc} can be
rewritten as
\begin{align}
  S = \frac{A}{4G}, \label{S:Wald}
\end{align}
which also agrees with the Bekenstein-Hawking entropy.  After all, we
obtained the Wald entropy expressed by the expected area law, which is
independent of the normalization of the Killing vector.

\section{Entropy from brick-wall method}
\label{sec:brick-wall}

In the Nariai black hole governed by the line
element~\eqref{metric:Nariai}, the black hole temperature is the same
with the cosmological temperature as seen from
Eqs.~\eqref{T:BH:Nariai} and~\eqref{T:H:Nariai}, which imply that the
net flux is in fact zero. Thus, the thermal equilibrium can be
realized in this special configuration, which is different from the
non-equilibrium SdS black hole.  In order to calculate the statistical
entropy in this thermal background~\cite{'tHooft:1984re}, we will
consider a quantum scalar field in a box surrounded by the two
horizons.  The Klein-Gordon equation for the scalar field is written
as
\begin{align}
  (\Box - m^2) \Phi = 0, \label{KG}
\end{align}
where $m$ is the mass of the scalar field.  By using the WKB
approximation with $\Phi \sim \exp[-i\omega \psi + i S(\chi, \theta,
\phi)]$ under the Nariai metric~\eqref{metric:Nariai}, the square
module of the momentum is obtained as
\begin{align}
  k^2 = g^{\mu\nu} k_\mu k_\nu = \Lambda \left( -
    \frac{\omega^2}{\sin^2\chi} + k_\chi^2 + k_\theta^2 +
    \frac{k_\phi^2}{\sin^2\theta} \right) = -m^2, \label{ksquare}
\end{align}
where $k_\chi = \partial S/\partial \chi$, $k_\theta = \partial
S/\partial \theta$, and $k_\phi = \partial S/\partial \phi$. Then, the
number of quantum states with the energy less than $\omega$ is
calculated as
\begin{align}
  n(\omega) &= \frac{1}{(2\pi)^3} \int_{V_p} d\chi d\theta d\phi
  dk_\chi dk_\theta dk_\phi \notag \\
  &= \frac{2}{3\pi} \int \frac{d\chi}{\sin^3\chi} \left(\omega^2 -
    \frac{m^2}{\Lambda} \sin^2\chi \right)^{3/2}, \label{n:omega}
\end{align}
where $V_p$ denotes the volume of the phase space satisfying $k^2 +
m^2 \le 0$. For simplicity, we take the massless limit of $m^2 =0$.
As seen from~\eqref{n:omega}, the number of states diverges at the
horizons of $\chi =0, \pi$, so that we need the UV cutoff at
$\chi=h_b$ and $\chi = \pi - h_c$. The UV cutoff parameters $h_b$ and
$h_c$ are assumed to be very small.  Then, the free energy is given by
\begin{align}
  F &= - \int d\omega \frac{n(\omega)}{e^{\beta\omega} - 1} \notag \\
  &= - \frac{2}{3\pi} \int_{h_b}^{\pi - h_c} \frac{d\chi}{\sin^3\chi}
  \int_0^\infty d\omega \frac{\omega^3}{e^{\beta\omega}-1} \notag \\
  &= - \frac{\pi^3}{45 \beta^4} \left[- \frac{\cos\chi}{\sin^2\chi} +
    \ln\left(\tan\frac{\chi}{2} \right) \right]_{h_b}^{\pi - h_c}
  \notag \\
  &= - \frac{\pi^3}{45 \beta^4} \left[\frac{1}{h_b^2} - \ln h_b +
    \frac{1}{h_c^2} - \ln h_c + O(h_b^0, h_c^0)
  \right]. \label{F:cutoff}
\end{align}
Then, the entropy becomes
\begin{align}
  S = \beta^2 \frac{\partial F}{\partial\beta} = \frac{4\pi^3}{45
    \beta^3} \left[\frac{1}{h_b^2} - \ln h_b + \frac{1}{h_c^2} - \ln
    h_c + O(h_b^0, h_c^0) \right]. \label{S:cutoff}
\end{align}
The proper lengths for the UV parameters are given by 
\begin{align}
  \bar{h}_b &= \int_0^{h_b} d\chi \sqrt{g_{\chi\chi}} =
  \frac{h_b}{\sqrt{\Lambda}}, \label{cutoff:bh:proper} \\
  \bar{h}_c &= \int_0^{\pi - h_c} d\chi \sqrt{g_{\chi\chi}} =
  \frac{h_c}{\sqrt{\Lambda}}, \label{cutoff:ch:proper}
\end{align}
which leads to $h_{b,c} = \sqrt{\Lambda} \, \bar{h}_{b,c}$.  Then,
Eq.~\eqref{S:cutoff} is written as
\begin{align}
  S = \frac{4\pi^3}{45 \beta^3} \left(\frac{1}{\Lambda \bar{h}_b^2} +
    \frac{1}{\Lambda \bar{h}_c^2} \right), \label{S:cutoff:proper}
\end{align}
within the leading order of $\bar{h}_{b,c}$.  

When we perform the WKB approximation with the line
element~\eqref{metric:Nariai}, the coordinate $\psi$ plays a role of
the time. The corresponding Killing vector is given by $\xi
= \partial/ \partial \psi$ and $\beta$ in Eq.~\eqref{S:cutoff:proper}
should be taken as the inverse of the Hawking
temperature~\eqref{T:H:Nariai}. Then, the entropy is obtained as
\begin{align}
  S = \frac{\ell_{\rm P}^2}{90\pi \bar{h}_b^2}
  \frac{c^3A_b}{4G\hbar} + \frac{\ell_{\rm P}^2}{90\pi
    \bar{h}_c^2}\frac{c^3A_c}{4G\hbar}, \label{S:dim}
\end{align}
where $\ell_{\rm P} \equiv \sqrt{G\hbar/c^3}$ is the Plank length.  If
the cutoff is chosen as $\bar{h}_{b,c} = \ell_{\rm P}/\sqrt{90\pi}$
like the case of the Schwarzschild black hole~\cite{'tHooft:1984re},
the entropy~\eqref{S:dim} is remarkably written as
\begin{align}
  S = \frac{c^3A}{4G\hbar}, \label{S:final}
\end{align}
where the total area is defined by $A = A_b + A_c $ for convenience.
Then, it agrees with one quarter of the horizon area of the
Bekenstein-Hawking entropy.

From the viewpoint of the renormalization~\cite{Demers:1995dq}, the
total entropy can be written as the sum of the Wald entropy
\eqref{S:Wald} and the quantum correction of Eq.~\eqref{S:dim}.  If we
consider the bare gravitational coupling constant in the classical
entropy, the divergent part can be easily absorbed in the
gravitational constant.

\section{Discussion}
\label{sec:discus}

By using the brick-wall method for the Nariai spacetime, we obtained
the Bekenstein-Hawking entropy which is proportional to the area of
the horizon. In the brick-wall method, $\beta$ was not the inverse of
the Bousso-Hawking temperature but the inverse of the Hawking
temperature.  The reason why is that the time is chosen as $\psi$ and
the standard form of the corresponding Killing vector is given by
$\partial/ \partial\psi$. If we consider the scaled time $\tilde{\psi}
= \psi/\sqrt{\Lambda}$, the Killing vector is given by $\xi
= \partial/\partial \tilde{\psi} = \sqrt{\Lambda} \partial /\partial
\psi$, which yields the Bousso-Hawking
temperature~\eqref{T:BH:Nariai}. Then, the WKB approximation in the
brick-wall method should be performed for the scalar field in the form
of $\Phi \sim \exp[-i\tilde{\omega} \tilde{\psi} + i
S(\chi,\theta,\phi)] = \exp[-i\tilde{\omega} \psi /\sqrt{\Lambda} + i
S(\chi,\theta,\phi)]$. This indicates that the energy in
Eq.~\eqref{ksquare} becomes $\tilde{\omega} = \omega \sqrt{\Lambda}$
and we can easily show that $\beta_H \omega = \beta_{\rm BH}
\tilde{\omega}$. In the calculation of the free
energy~\eqref{F:cutoff}, $\omega$ in the integrand should be replaced
by $\tilde{\omega} /\sqrt{\Lambda}$ and the integration should be
performed for $\tilde{\omega}$. Then, we can obtain the same entropy
with Eq.~\eqref{S:dim} based on the Bousso-Hawking
temperature. Therefore, the entropy is always written as the area law
of the Wald entropy, whereas the temperature and the energy depend on
the choice of the time, that is, the normalization of the timelike
Killing vector.

The final comment is in order. As for the Bousso-Hawking temperature,
it can be regarded as a Tolman temperature~\cite{Tolman:1930zza}.  It
was defined at the vanishing surface gravity where it is the
counterpart of the asymptotically Minkowski space in the
asymptotically flat black holes. The Bousso-Hawking temperature can be
derived from the definition of the Tolman temperature of $T_{\rm{loc}}
= T_{H}/\sqrt{g_{\psi\psi}} = \sqrt{\Lambda} / (2\pi \sin\chi)$ where
$T_{H} = 1/(2\pi)$.  If we move the observer, for instance, to the
black hole horizon of $\chi=0$ or to the cosmological horizon of
$\chi=\pi$, the temperature goes to infinity. In particular, at the
middle point of $\chi = \frac{\pi}{2}$, it produces the Bousso-Hawking
temperature. So the Bousso-Hawking normalization of Killing vector is
compatible with the Tolman temperature.  So, we can identify the
Bousso-Hawking temperature with the Tolman temperature at the
reference point.
  

\begin{acknowledgments}
  We would like to thank Y.~Kim for exciting discussion.  M.~Eune was
  supported by National Research Foundation of Korea Grant funded by
  the Korean Government (Ministry of Education, Science and
  Technology) (NRF-2010-359-C00007). W.~Kim are supported by the
  National Research Foundation of Korea(NRF) grant funded by the Korea
  government(MEST) through the Center for Quantum Spacetime(CQUeST) of
  Sogang University with grant number 2005-0049409, and the Basic
  Science Research Program through the National Research Foundation of
  Korea(NRF) funded by the Ministry of Education, Science and
  Technology(2012-0002880).
\end{acknowledgments}


\bibliographystyle{apsrev} 
\bibliography{references}

\begin{thebibliography}{43}
\expandafter\ifx\csname natexlab\endcsname\relax\def\natexlab#1{#1}\fi
\expandafter\ifx\csname bibnamefont\endcsname\relax
  \def\bibnamefont#1{#1}\fi
\expandafter\ifx\csname bibfnamefont\endcsname\relax
  \def\bibfnamefont#1{#1}\fi
\expandafter\ifx\csname citenamefont\endcsname\relax
  \def\citenamefont#1{#1}\fi
\expandafter\ifx\csname url\endcsname\relax
  \def\url#1{\texttt{#1}}\fi
\expandafter\ifx\csname urlprefix\endcsname\relax\def\urlprefix{URL }\fi
\providecommand{\bibinfo}[2]{#2}
\providecommand{\eprint}[2][]{\url{#2}}

\bibitem[{\citenamefont{Bekenstein}(1973)}]{Bekenstein:1973ur}
\bibinfo{author}{\bibfnamefont{J.~D.} \bibnamefont{Bekenstein}},
  \bibinfo{journal}{Phys. Rev.} \textbf{\bibinfo{volume}{D7}},
  \bibinfo{pages}{2333} (\bibinfo{year}{1973}).

\bibitem[{\citenamefont{Hawking}(1975)}]{Hawking:1974sw}
\bibinfo{author}{\bibfnamefont{S.}~\bibnamefont{Hawking}},
  \bibinfo{journal}{Commun. Math. Phys.} \textbf{\bibinfo{volume}{43}},
  \bibinfo{pages}{199} (\bibinfo{year}{1975}).

\bibitem[{\citenamefont{'t~Hooft}(1985)}]{'tHooft:1984re}
\bibinfo{author}{\bibfnamefont{G.}~\bibnamefont{'t~Hooft}},
  \bibinfo{journal}{Nucl. Phys.} \textbf{\bibinfo{volume}{B256}},
  \bibinfo{pages}{727} (\bibinfo{year}{1985}).

\bibitem[{\citenamefont{Mann et~al.}(1992)\citenamefont{Mann, Tarasov, and
  Zelnikov}}]{Mann:1990fk}
\bibinfo{author}{\bibfnamefont{R.~B.} \bibnamefont{Mann}},
  \bibinfo{author}{\bibfnamefont{L.}~\bibnamefont{Tarasov}}, \bibnamefont{and}
  \bibinfo{author}{\bibfnamefont{A.}~\bibnamefont{Zelnikov}},
  \bibinfo{journal}{Class. Quant. Grav.} \textbf{\bibinfo{volume}{9}},
  \bibinfo{pages}{1487} (\bibinfo{year}{1992}).

\bibitem[{\citenamefont{Ghosh and Mitra}(1994)}]{Ghosh:1994wb}
\bibinfo{author}{\bibfnamefont{A.}~\bibnamefont{Ghosh}} \bibnamefont{and}
  \bibinfo{author}{\bibfnamefont{P.}~\bibnamefont{Mitra}},
  \bibinfo{journal}{Phys. Rev. Lett.} \textbf{\bibinfo{volume}{73}},
  \bibinfo{pages}{2521} (\bibinfo{year}{1994}), \eprint{hep-th/9406210}.

\bibitem[{\citenamefont{Ghosh and Mitra}(1995)}]{Ghosh:1994mm}
\bibinfo{author}{\bibfnamefont{A.}~\bibnamefont{Ghosh}} \bibnamefont{and}
  \bibinfo{author}{\bibfnamefont{P.}~\bibnamefont{Mitra}},
  \bibinfo{journal}{Phys. Lett.} \textbf{\bibinfo{volume}{B357}},
  \bibinfo{pages}{295} (\bibinfo{year}{1995}), \eprint{hep-th/9411128}.

\bibitem[{\citenamefont{Solodukhin}(1996)}]{Solodukhin:1996vx}
\bibinfo{author}{\bibfnamefont{S.~N.} \bibnamefont{Solodukhin}},
  \bibinfo{journal}{Phys. Rev.} \textbf{\bibinfo{volume}{D54}},
  \bibinfo{pages}{3900} (\bibinfo{year}{1996}), \eprint{hep-th/9601154}.

\bibitem[{\citenamefont{Ho and Kang}(1998)}]{Ho:1998du}
\bibinfo{author}{\bibfnamefont{J.-w.} \bibnamefont{Ho}} \bibnamefont{and}
  \bibinfo{author}{\bibfnamefont{G.}~\bibnamefont{Kang}},
  \bibinfo{journal}{Phys. Lett.} \textbf{\bibinfo{volume}{B445}},
  \bibinfo{pages}{27} (\bibinfo{year}{1998}), \eprint{gr-qc/9806118}.

\bibitem[{\citenamefont{Li and Zhao}(2000)}]{Li:2000rk}
\bibinfo{author}{\bibfnamefont{X.}~\bibnamefont{Li}} \bibnamefont{and}
  \bibinfo{author}{\bibfnamefont{Z.}~\bibnamefont{Zhao}},
  \bibinfo{journal}{Phys. Rev.} \textbf{\bibinfo{volume}{D62}},
  \bibinfo{pages}{104001} (\bibinfo{year}{2000}).

\bibitem[{\citenamefont{Winstanley}(2001)}]{Winstanley:2000in}
\bibinfo{author}{\bibfnamefont{E.}~\bibnamefont{Winstanley}},
  \bibinfo{journal}{Phys. Rev.} \textbf{\bibinfo{volume}{D63}},
  \bibinfo{pages}{084013} (\bibinfo{year}{2001}), \eprint{hep-th/0011176}.

\bibitem[{\citenamefont{Li}(2002)}]{Li:2002xb}
\bibinfo{author}{\bibfnamefont{X.}~\bibnamefont{Li}}, \bibinfo{journal}{Phys.
  Lett.} \textbf{\bibinfo{volume}{B540}}, \bibinfo{pages}{9}
  (\bibinfo{year}{2002}), \eprint{gr-qc/0204029}.

\bibitem[{\citenamefont{Kim et~al.}(2006)\citenamefont{Kim, Son, Yoon, and
  Park}}]{Kim:2005pb}
\bibinfo{author}{\bibfnamefont{W.~T.} \bibnamefont{Kim}},
  \bibinfo{author}{\bibfnamefont{E.~J.} \bibnamefont{Son}},
  \bibinfo{author}{\bibfnamefont{M.~S.} \bibnamefont{Yoon}}, \bibnamefont{and}
  \bibinfo{author}{\bibfnamefont{Y.-J.} \bibnamefont{Park}},
  \bibinfo{journal}{J. Korean Phys. Soc.} \textbf{\bibinfo{volume}{49}},
  \bibinfo{pages}{15} (\bibinfo{year}{2006}), \eprint{gr-qc/0504127}.

\bibitem[{\citenamefont{Yoon et~al.}(2007)\citenamefont{Yoon, Ha, and
  Kim}}]{Yoon:2007aj}
\bibinfo{author}{\bibfnamefont{M.}~\bibnamefont{Yoon}},
  \bibinfo{author}{\bibfnamefont{J.}~\bibnamefont{Ha}}, \bibnamefont{and}
  \bibinfo{author}{\bibfnamefont{W.}~\bibnamefont{Kim}},
  \bibinfo{journal}{Phys. Rev.} \textbf{\bibinfo{volume}{D76}},
  \bibinfo{pages}{047501} (\bibinfo{year}{2007}), \eprint{0706.0364}.

\bibitem[{\citenamefont{Kim and Park}(2007)}]{Kim:2007if}
\bibinfo{author}{\bibfnamefont{Y.-W.} \bibnamefont{Kim}} \bibnamefont{and}
  \bibinfo{author}{\bibfnamefont{Y.-J.} \bibnamefont{Park}},
  \bibinfo{journal}{Phys. Lett.} \textbf{\bibinfo{volume}{B655}},
  \bibinfo{pages}{172} (\bibinfo{year}{2007}), \eprint{0707.2128}.

\bibitem[{\citenamefont{Kim et~al.}(2007)\citenamefont{Kim, Kim, and
  Park}}]{Kim:2007nh}
\bibinfo{author}{\bibfnamefont{W.}~\bibnamefont{Kim}},
  \bibinfo{author}{\bibfnamefont{Y.-W.} \bibnamefont{Kim}}, \bibnamefont{and}
  \bibinfo{author}{\bibfnamefont{Y.-J.} \bibnamefont{Park}},
  \bibinfo{journal}{Phys. Rev.} \textbf{\bibinfo{volume}{D75}},
  \bibinfo{pages}{127501} (\bibinfo{year}{2007}), \eprint{gr-qc/0702018}.

\bibitem[{\citenamefont{Ren et~al.}(2008)\citenamefont{Ren, Li-Chun, Huai-Fan,
  and Yue-Qin}}]{Ren:2008rv}
\bibinfo{author}{\bibfnamefont{Z.}~\bibnamefont{Ren}},
  \bibinfo{author}{\bibfnamefont{Z.}~\bibnamefont{Li-Chun}},
  \bibinfo{author}{\bibfnamefont{L.}~\bibnamefont{Huai-Fan}}, \bibnamefont{and}
  \bibinfo{author}{\bibfnamefont{W.}~\bibnamefont{Yue-Qin}},
  \bibinfo{journal}{Int. J. Theor. Phys.} \textbf{\bibinfo{volume}{47}},
  \bibinfo{pages}{3083} (\bibinfo{year}{2008}), \eprint{0801.4118}.

\bibitem[{\citenamefont{Liu et~al.}(2008)\citenamefont{Liu, Gui, and
  Liu}}]{Liu:2008bb}
\bibinfo{author}{\bibfnamefont{M.}~\bibnamefont{Liu}},
  \bibinfo{author}{\bibfnamefont{Y.}~\bibnamefont{Gui}}, \bibnamefont{and}
  \bibinfo{author}{\bibfnamefont{H.}~\bibnamefont{Liu}},
  \bibinfo{journal}{Phys. Rev.} \textbf{\bibinfo{volume}{D78}},
  \bibinfo{pages}{124003} (\bibinfo{year}{2008}), \eprint{0812.0864}.

\bibitem[{\citenamefont{Sarkar et~al.}(2008)\citenamefont{Sarkar,
  Shankaranarayanan, and Sriramkumar}}]{Sarkar:2007uz}
\bibinfo{author}{\bibfnamefont{S.}~\bibnamefont{Sarkar}},
  \bibinfo{author}{\bibfnamefont{S.}~\bibnamefont{Shankaranarayanan}},
  \bibnamefont{and}
  \bibinfo{author}{\bibfnamefont{L.}~\bibnamefont{Sriramkumar}},
  \bibinfo{journal}{Phys. Rev.} \textbf{\bibinfo{volume}{D78}},
  \bibinfo{pages}{024003} (\bibinfo{year}{2008}), \eprint{0710.2013}.

\bibitem[{\citenamefont{Ghosh}(2009)}]{Ghosh:2009xg}
\bibinfo{author}{\bibfnamefont{K.}~\bibnamefont{Ghosh}},
  \bibinfo{journal}{Nucl. Phys.} \textbf{\bibinfo{volume}{B814}},
  \bibinfo{pages}{212} (\bibinfo{year}{2009}), \eprint{0902.1601}.

\bibitem[{\citenamefont{Kim and Son}(2009)}]{Kim:2008bf}
\bibinfo{author}{\bibfnamefont{W.}~\bibnamefont{Kim}} \bibnamefont{and}
  \bibinfo{author}{\bibfnamefont{E.~J.} \bibnamefont{Son}},
  \bibinfo{journal}{Phys. Lett.} \textbf{\bibinfo{volume}{B673}},
  \bibinfo{pages}{90} (\bibinfo{year}{2009}), \eprint{0812.0876}.

\bibitem[{\citenamefont{Chang-Young et~al.}(2009)\citenamefont{Chang-Young,
  Lee, and Yoon}}]{ChangYoung:2008pw}
\bibinfo{author}{\bibfnamefont{E.}~\bibnamefont{Chang-Young}},
  \bibinfo{author}{\bibfnamefont{D.}~\bibnamefont{Lee}}, \bibnamefont{and}
  \bibinfo{author}{\bibfnamefont{M.}~\bibnamefont{Yoon}},
  \bibinfo{journal}{Class. Quant. Grav.} \textbf{\bibinfo{volume}{26}},
  \bibinfo{pages}{155011} (\bibinfo{year}{2009}), \eprint{0811.3294}.

\bibitem[{\citenamefont{Eune and Kim}(2010)}]{Eune:2010kx}
\bibinfo{author}{\bibfnamefont{M.}~\bibnamefont{Eune}} \bibnamefont{and}
  \bibinfo{author}{\bibfnamefont{W.}~\bibnamefont{Kim}},
  \bibinfo{journal}{Phys. Rev.} \textbf{\bibinfo{volume}{D82}},
  \bibinfo{pages}{124048} (\bibinfo{year}{2010}), \eprint{1007.1824}.

\bibitem[{\citenamefont{Kay and Ortiz}(2011)}]{Kay:2011np}
\bibinfo{author}{\bibfnamefont{B.~S.} \bibnamefont{Kay}} \bibnamefont{and}
  \bibinfo{author}{\bibfnamefont{L.}~\bibnamefont{Ortiz}}
  (\bibinfo{year}{2011}), \eprint{1111.6429}.

\bibitem[{\citenamefont{Son et~al.}(2012)\citenamefont{Son, Eune, and
  Kim}}]{Son:2011ii}
\bibinfo{author}{\bibfnamefont{E.~J.} \bibnamefont{Son}},
  \bibinfo{author}{\bibfnamefont{M.}~\bibnamefont{Eune}}, \bibnamefont{and}
  \bibinfo{author}{\bibfnamefont{W.}~\bibnamefont{Kim}},
  \bibinfo{journal}{Phys. Lett.} \textbf{\bibinfo{volume}{B706}},
  \bibinfo{pages}{447} (\bibinfo{year}{2012}), \eprint{1109.5486}.

\bibitem[{\citenamefont{Zhao et~al.}(2011)\citenamefont{Zhao, He, and
  Tang}}]{Zhao:2011zzc}
\bibinfo{author}{\bibfnamefont{F.}~\bibnamefont{Zhao}},
  \bibinfo{author}{\bibfnamefont{F.}~\bibnamefont{He}}, \bibnamefont{and}
  \bibinfo{author}{\bibfnamefont{J.}~\bibnamefont{Tang}},
  \bibinfo{journal}{Phys. Rev.} \textbf{\bibinfo{volume}{D84}},
  \bibinfo{pages}{104007} (\bibinfo{year}{2011}).

\bibitem[{\citenamefont{Cao and He}(2012)}]{Cao:2012zzb}
\bibinfo{author}{\bibfnamefont{F.}~\bibnamefont{Cao}} \bibnamefont{and}
  \bibinfo{author}{\bibfnamefont{F.}~\bibnamefont{He}}, \bibinfo{journal}{Int.
  J. Theor. Phys.} \textbf{\bibinfo{volume}{51}}, \bibinfo{pages}{536}
  (\bibinfo{year}{2012}).

\bibitem[{\citenamefont{Wald}(1993)}]{Wald:1993nt}
\bibinfo{author}{\bibfnamefont{R.~M.} \bibnamefont{Wald}},
  \bibinfo{journal}{Phys. Rev.} \textbf{\bibinfo{volume}{D48}},
  \bibinfo{pages}{3427} (\bibinfo{year}{1993}), \eprint{gr-qc/9307038}.

\bibitem[{\citenamefont{Iyer and Wald}(1995)}]{Iyer:1995kg}
\bibinfo{author}{\bibfnamefont{V.}~\bibnamefont{Iyer}} \bibnamefont{and}
  \bibinfo{author}{\bibfnamefont{R.~M.} \bibnamefont{Wald}},
  \bibinfo{journal}{Phys. Rev.} \textbf{\bibinfo{volume}{D52}},
  \bibinfo{pages}{4430} (\bibinfo{year}{1995}), \eprint{gr-qc/9503052}.

\bibitem[{\citenamefont{Jacobson et~al.}(1994)\citenamefont{Jacobson, Kang, and
  Myers}}]{Jacobson:1993vj}
\bibinfo{author}{\bibfnamefont{T.}~\bibnamefont{Jacobson}},
  \bibinfo{author}{\bibfnamefont{G.}~\bibnamefont{Kang}}, \bibnamefont{and}
  \bibinfo{author}{\bibfnamefont{R.~C.} \bibnamefont{Myers}},
  \bibinfo{journal}{Phys. Rev.} \textbf{\bibinfo{volume}{D49}},
  \bibinfo{pages}{6587} (\bibinfo{year}{1994}), \eprint{gr-qc/9312023}.

\bibitem[{\citenamefont{Jacobson and Myers}(1993)}]{Jacobson:1993xs}
\bibinfo{author}{\bibfnamefont{T.}~\bibnamefont{Jacobson}} \bibnamefont{and}
  \bibinfo{author}{\bibfnamefont{R.~C.} \bibnamefont{Myers}},
  \bibinfo{journal}{Phys. Rev.Lett.} \textbf{\bibinfo{volume}{70}},
  \bibinfo{pages}{3684} (\bibinfo{year}{1993}), \eprint{hep-th/9305016}.

\bibitem[{\citenamefont{Lopes~Cardoso et~al.}(2000)\citenamefont{Lopes~Cardoso,
  de~Wit, and Mohaupt}}]{LopesCardoso:1999cv}
\bibinfo{author}{\bibfnamefont{G.}~\bibnamefont{Lopes~Cardoso}},
  \bibinfo{author}{\bibfnamefont{B.}~\bibnamefont{de~Wit}}, \bibnamefont{and}
  \bibinfo{author}{\bibfnamefont{T.}~\bibnamefont{Mohaupt}},
  \bibinfo{journal}{Fortsch. Phys.} \textbf{\bibinfo{volume}{48}},
  \bibinfo{pages}{49} (\bibinfo{year}{2000}), \eprint{hep-th/9904005}.

\bibitem[{\citenamefont{Parikh and Sarkar}(2009)}]{Parikh:2009qs}
\bibinfo{author}{\bibfnamefont{M.~K.} \bibnamefont{Parikh}} \bibnamefont{and}
  \bibinfo{author}{\bibfnamefont{S.}~\bibnamefont{Sarkar}}
  (\bibinfo{year}{2009}), \eprint{0903.1176}.

\bibitem[{\citenamefont{Parikh}(2011)}]{Parikh:2009js}
\bibinfo{author}{\bibfnamefont{M.~K.} \bibnamefont{Parikh}},
  \bibinfo{journal}{Phys.Rev.} \textbf{\bibinfo{volume}{D84}},
  \bibinfo{pages}{044048} (\bibinfo{year}{2011}), \eprint{0909.3307}.

\bibitem[{\citenamefont{Zhao et~al.}(2001)\citenamefont{Zhao, Zhang, and
  Zhang}}]{Zhao:2001pf}
\bibinfo{author}{\bibfnamefont{R.}~\bibnamefont{Zhao}},
  \bibinfo{author}{\bibfnamefont{J.-F.} \bibnamefont{Zhang}}, \bibnamefont{and}
  \bibinfo{author}{\bibfnamefont{L.-C.} \bibnamefont{Zhang}},
  \bibinfo{journal}{Mod. Phys. Lett.} \textbf{\bibinfo{volume}{A16}},
  \bibinfo{pages}{719} (\bibinfo{year}{2001}).

\bibitem[{\citenamefont{Wu et~al.}(2001)\citenamefont{Wu, Zhang, and
  Zhao}}]{Wu:2001uy}
\bibinfo{author}{\bibfnamefont{Y.-Q.} \bibnamefont{Wu}},
  \bibinfo{author}{\bibfnamefont{L.-C.} \bibnamefont{Zhang}}, \bibnamefont{and}
  \bibinfo{author}{\bibfnamefont{R.}~\bibnamefont{Zhao}},
  \bibinfo{journal}{Int. J. Theor. Phys.} \textbf{\bibinfo{volume}{40}},
  \bibinfo{pages}{1001} (\bibinfo{year}{2001}).

\bibitem[{\citenamefont{Ding and Jing}(2007)}]{Ding:2007ur}
\bibinfo{author}{\bibfnamefont{C.}~\bibnamefont{Ding}} \bibnamefont{and}
  \bibinfo{author}{\bibfnamefont{J.}~\bibnamefont{Jing}},
  \bibinfo{journal}{JHEP} \textbf{\bibinfo{volume}{0709}}, \bibinfo{pages}{067}
  (\bibinfo{year}{2007}), \eprint{0709.1186}.

\bibitem[{\citenamefont{Wu and Yan}(2004)}]{Wu:2003qc}
\bibinfo{author}{\bibfnamefont{S.-Q.} \bibnamefont{Wu}} \bibnamefont{and}
  \bibinfo{author}{\bibfnamefont{M.-L.} \bibnamefont{Yan}},
  \bibinfo{journal}{Phys. Rev.} \textbf{\bibinfo{volume}{D69}},
  \bibinfo{pages}{044019} (\bibinfo{year}{2004}), \eprint{gr-qc/0303076}.

\bibitem[{\citenamefont{Ginsparg and Perry}(1983)}]{Ginsparg:1982rs}
\bibinfo{author}{\bibfnamefont{P.~H.} \bibnamefont{Ginsparg}} \bibnamefont{and}
  \bibinfo{author}{\bibfnamefont{M.~J.} \bibnamefont{Perry}},
  \bibinfo{journal}{Nucl. Phys.} \textbf{\bibinfo{volume}{B222}},
  \bibinfo{pages}{245} (\bibinfo{year}{1983}).

\bibitem[{\citenamefont{Bousso and Hawking}(1996)}]{Bousso:1996au}
\bibinfo{author}{\bibfnamefont{R.}~\bibnamefont{Bousso}} \bibnamefont{and}
  \bibinfo{author}{\bibfnamefont{S.~W.} \bibnamefont{Hawking}},
  \bibinfo{journal}{Phys. Rev.} \textbf{\bibinfo{volume}{D54}},
  \bibinfo{pages}{6312} (\bibinfo{year}{1996}), \eprint{gr-qc/9606052}.

\bibitem[{\citenamefont{Bousso and Hawking}(1998)}]{Bousso:1997wi}
\bibinfo{author}{\bibfnamefont{R.}~\bibnamefont{Bousso}} \bibnamefont{and}
  \bibinfo{author}{\bibfnamefont{S.~W.} \bibnamefont{Hawking}},
  \bibinfo{journal}{Phys. Rev.} \textbf{\bibinfo{volume}{D57}},
  \bibinfo{pages}{2436} (\bibinfo{year}{1998}), \eprint{hep-th/9709224}.

\bibitem[{\citenamefont{Dias and Lemos}(2003)}]{Dias:2003up}
\bibinfo{author}{\bibfnamefont{O.~J.} \bibnamefont{Dias}} \bibnamefont{and}
  \bibinfo{author}{\bibfnamefont{J.~P.} \bibnamefont{Lemos}},
  \bibinfo{journal}{Phys. Rev.} \textbf{\bibinfo{volume}{D68}},
  \bibinfo{pages}{104010} (\bibinfo{year}{2003}), \eprint{hep-th/0306194}.

\bibitem[{\citenamefont{Demers et~al.}(1995)\citenamefont{Demers, Lafrance, and
  Myers}}]{Demers:1995dq}
\bibinfo{author}{\bibfnamefont{J.-G.} \bibnamefont{Demers}},
  \bibinfo{author}{\bibfnamefont{R.}~\bibnamefont{Lafrance}}, \bibnamefont{and}
  \bibinfo{author}{\bibfnamefont{R.~C.} \bibnamefont{Myers}},
  \bibinfo{journal}{Phys. Rev.} \textbf{\bibinfo{volume}{D52}},
  \bibinfo{pages}{2245} (\bibinfo{year}{1995}), \eprint{gr-qc/9503003}.

\bibitem[{\citenamefont{Tolman}(1930)}]{Tolman:1930zza}
\bibinfo{author}{\bibfnamefont{R.~C.} \bibnamefont{Tolman}},
  \bibinfo{journal}{Phys. Rev.} \textbf{\bibinfo{volume}{35}},
  \bibinfo{pages}{904} (\bibinfo{year}{1930}).

\end{thebibliography}


\end{document}